\documentstyle[epsfig,12pt,aasms4]{article}
\def\sgra{SGR~1806--20}
\def\sgrb{SGR~1900+14}

\def\sgrd{SGR~1627--41}

\def\n49{N~49}
\def\w31{W~31}
\def\ga{G~10.0--0.3}

\def\gg{G~337.0--0.1}
\def\sax{{\em BeppoSAX}}
\def\batse{{\em BATSE}}
\def\asca{{\em ASCA}}
\def\xte{{\em RXTE}}
\def\ulysse{{\em Ulysses}}

\def\co{$^{12}$CO(J=1--0)}
\def\kms{\mbox{km s$^{-1}$}}
\def\av{A$_{\rm v}$}
\def\sm{\mbox{M$_\odot$}}

\newcommand{\hi}{\mbox{H{\sc i}}}
\newcommand{\hii}{\mbox{H{\sc~ii}}}

\lefthead{Corbel et al.}
\righthead{The distance to \sgrd}
\slugcomment{Accepted for publication in ApJ Letters}

\begin{document}

\title{The Distance to the Soft Gamma Repeater \sgrd.
}
\author{S. Corbel\altaffilmark{1}, C. Chapuis\altaffilmark{1}, T.M. Dame\altaffilmark{2} and P. Durouchoux\altaffilmark{1}}
\altaffiltext{1}{Service d'Astrophysique, CEA  Saclay, 91191 Gif sur Yvette Cedex, France}
\altaffiltext{2}{Harvard Smithsonian Center for Astrophysics, 60 Garden Street, Cambridge, MA 02138, USA}
\authoremail{corbel@discovery.saclay.cea.fr}

\begin{abstract}

We report millimeter observations of the line of sight to the recently discovered Soft Gamma Repeater,
\sgrd, which has been tentatively associated with the supernova remnant SNR \gg.
Among the eight molecular clouds along the line of sight to \sgrd, we show that SNR \gg\ is probably interacting 
with one of the most massive giant molecular clouds (GMC) in the Galaxy, at a distance of 11 kpc from the sun. 
Based on the high extinction to the persistent X-ray 
counterpart of \sgrd, we present evidence for an association of this new SGR with the SNR \gg; they both
appear to be located on the near side of the GMC. 
This is the second SGR located near an extraordinarily massive GMC. We suggest that \sgrd\  is 
a neutron star with a high transverse velocity ($\sim$ 1,000 \kms) escaping the young ($\sim$ 5,000 years) 
supernova remnant \gg. 

\end{abstract}


\section{Introduction}

Soft Gamma Repeaters (SGRs) are believed to be young neutron stars of high dipolar magnetic 
field, located in or near their associated supernova remnants (SNRs).
A key observation regarding the nature of these objects was the simultaneous detection by \asca\ and \batse\ 
of a burst from \sgra\ (\cite{kou94,mur94}), which allowed an accurate determination of the burst position. 
Subsequent observations revealed a persistent X-ray counterpart,
located near the center of the young plerionic SNR \ga\ (\cite{mur94,son94,kul94}).
Further insights into the nature of SGRs came from the detection by \xte\ of a very slow pulsation (7.47 s) 
with a very high spin down from the persistent X-ray counterpart of \sgra, implying an extremely strong
dipolar magnetic field (8 $\times$ 10$^{14}$ Gauss) for the neutron star (\cite{kou98a}). 
This results provided the first strong observational support for the magnetar model of Thomson \& Duncan (1995), in which
SGRs are young and strongly magnetized neutron stars. \sgra\ is associated with a giant molecular cloud (GMC) 
at a distance of 14.5 kpc from the sun (\cite{cor97}).
In 1998, an \asca\ observation of \sgrb\ during a bursting phase (\cite{mur99,hur99a}), provided an improved position,
and revealed a persistent X-ray counterpart, similar to that of \sgra.
\asca, \xte\ and \sax\ (\cite{hur99a,kou99,woo99a}) detected slow pulsation (5.16 s) from the persistent 
X-ray counterpart
with a very high spin down, giving a dipolar magnetic field of $\sim$ 5 $\times$ 10$^{14}$ Gauss, again 
supporting the magnetar model.

\sgrd\ was discovered with \batse\ (\cite{kou98b}). Its accurate error box, combining measurements from \ulysse, 
{\em Konus-Wind}, \xte\ and \batse\ (\cite{hur99b,smi99}) contains the SNR \gg\ (\cite{woo98}).
Observations with \sax\ revealed an X-ray source within the SNR, SAX~J1635.8--4736, with a non thermal 
X-ray spectrum (\cite{woo99b}) and a possible pulsation at 6.41 s. 
Based on the similar properties with other SGRs, it is very likely that they have detected 
the persistent X-ray counterpart of \sgrd. 

The very high interstellar absorption for \sgrd\ (Woods et al. 1999b) prompted us to perform millimeter 
observations in order to determine the source of this extinction. 
Here, we report observations carried out with the SEST telescope, allowing the detection of 8 distinct molecular clouds 
along the line of sight to \sgrd.
We show that \sgrd\ and the SNR \gg\ are both associated with an exceptionally massive GMC at a distance of 11 kpc.
We also show that the total extinction along the 11 kpc to the SGR, derived from CO and 21 cm observations, is
consistent with the foreground extinction derived for the X-ray counterpart.
We then discuss the consequences of this distance on the nature of \sgrd.

\section{Observations}

The observations were carried out with the 15 m Swedish -- ESO Submillimeter Telescope (SEST) at La Silla, Chile, 
on 1999 May 20. We observed the central position reported by Woods et al. (1999b).
We simultaneously observed two transitions: $^{12}$CO(J=1--0) and $^{12}$CO(J=2--1), at 115.27 and 230.54 GHz 
respectively; the J=2--1 line is shown in Figure 1 (solid line).
The FWHM beamwidth of SEST is 45\arcsec\ at 115.27 GHz and 23\arcsec\ at 230.54 GHz. 
We observed in position switching mode, after checking in frequency switching mode that our OFF position 
($\alpha$(2000) = 16h30m02.0s, $\delta$(2000) = --46\arcdeg 23\arcmin 32\arcsec) 
was free of emission. The back end was an acousto-optical spectrometer with a frequency bandwidth 
of 1 GHz and a velocity resolution of 1.8 \kms\ at 115.27 GHz and 0.9 \kms\ at 230.54 GHz. 
The system was calibrated with the chopper wheel
method; regular observations of calibration sources show that the SEST calibration is stable to within a few percent.
The system temperature during the observations was $\sim$ 284 K at 115.27 GHz and $\sim$ 197 K at 230.54 GHz. 
We obtained the 21 cm \hi\ spectra (dotted line in Figure 1) from the survey of neutral hydrogen in the southern Galaxy, 
observed by Kerr et al. (1986). The spectrum used for this study is an average of the two observed spectra 
closest to \sgrd, at l = 337\arcdeg, b = 0\arcdeg\ and --0.25\arcdeg\ (\sgrd\ is at l= 337.0\arcdeg\ 
and b = --0.1\arcdeg).

\section{Results}

As the velocity resolution is slightly better in the $^{12}$CO(J=2--1) spectrum (Figure 1) than the $^{12}$CO(J=1--0), 
we have used it to separate the various molecular clouds along the line of sight to \sgrd. 
The extinction has been derived from the $^{12}$CO(J=1--0) spectrum. 
A total of 8 clouds can be identified. We label each of them with the acronym MC and its associated velocity in Figure 1. 
It is well known that for gas in the inner Galaxy, there are two possible distances corresponding to each radial velocity. 
The parameters associated with each cloud are displayed in Table 1, using a circular model for the Galactic rotation 
(\cite{fic89}). 

As we are primarily interested in the distribution of extinction along the line of sight, 
we have estimated the atomic hydrogen column density, taking into account opacity effect, by using the following relation 
(\cite{roh96}):
$\frac{\mathcal{N}\mathrm{(H\ I)}}{\mathrm{atome\ cm}^{-2}} = - 1.82 \times 10^{18} \left( \frac{T_S}{\mathrm{K}} \right)
\int \ln \left( 1 - \frac{T_b(v)}{T_S - T_{bg}} \right) \frac{dv}{\mathrm{km~s}^{-1}}$,
and taking the usual assumption of T$_S$  = 125 K for the spin temperature and T$_{bg}$ = 2.7 K for the 
cosmic background temperature; T$_b$ is the 21 cm line brightness temperature. 
Molecular hydrogen column density, N(H$_2$), was derived from velocity integrated \co\ intensity, W(CO), using
a conversion factor, X$_{CO}$ $\equiv$  N(H$_2$)/W(CO), of (1.9 $\pm$ 0.2) $\times$ 10$^{20}$ mols. cm$^{-2}$/(K \kms).
This value was determined by comparing the $\gamma$-ray, \hi\ and $^{12}$CO(J=1--0) emissions from the Galaxy (Strong \& 
Mattox, 1996). The $^{12}$CO(J=1--0) transition has been shown to be the best tracer of 
molecular cloud masses (\cite{com91}), and three different methods have led to a similar value 
for X$_{CO}$ (\cite{sol91}).
The total hydrogen column density, N(H),  is then N(H) = N(\hi) + 2 N(H$_2$). We can estimate the total extinction 
using the following relation (\cite{pre95}):
\av = (5.6 $\pm$ 0.1 ) $\times$ 10$^{-22}\times \mathcal{N}$(H).
Table 2 gives the extinction for each of the molecular clouds.

\section{The supernova remnant \gg}

In the MOST supernova remnant catalog (Whiteoak \& Green 1996), \gg\ has a non thermal spectrum and a peculiar morphology, 
with several intensity enhancements superposed on a plateau of emission. It is a member of the CTB 33 complex,
which also includes the \hii\ regions G~337.1--0.2 and G~336.8+0.0. 
Using high resolution radio observations, Sarma et al. (1997) 
clearly resolved the \gg\ area into three components: another \hii\ region, the supernova remnant \gg\ and an edge
source. Frail et al. (1996) detected a maser OH emission line from the SNR, \gg, at --71.3 \kms. As mentioned 
by Sarma et al. (1997), this OH line is in agreement with the velocity (--73 to --75 \kms) of the recombination lines from the 
\hii\ regions of the CTB~33 complex. The \hi\ absorption profiles (\cite{sar97}) toward the members of the CTB~33 
complex (including the SNR \gg) show absorption features up to the tangent point, therefore ruling out the near
distance. Based on the rotation curve model of Fich et al. (1989),
a distance of 11.0 $\pm$ 0.3 kpc is deduced for the CTB~33 complex, including the SNR \gg\ (\cite{sar97}).

In the $^{12}$CO(J=2--1) spectrum toward \sgrd\ (Figure 1), which encompasses the direction of \gg, a GMC is 
detected at --70.9 \kms, close to the velocity of the components of the CTB~33 complex. 
In order to image the molecular emission in the vicinity of MC--71, we used
the CO survey of Bronfman et al. (1989).  The CO map in Figure 2 is
integrated over the velocity range --80 to --40 \kms, which includes all of
the emission from  MC--71 as well as that from an adjacent cloud at a
velocity of --56 \kms\ (MC--56) which may be related (see below).

The coincidence of MC--71 in direction and velocity with both the \hii\ 
regions G~337.1--0.2 and G~336.8+0.0 and the SNR \gg\ strongly suggests that MC--71 is the
progenitor of the entire CTB 33 complex, which lies at the far kinematic
distance of 11 kpc.  The radius-linewidth relation for GMCs (Dame et al.
1986) also favors the far kinematic distance for MC--71:  this GMC has an
exceptionally large composite linewidth of $\sim$ 19 \kms\ but a relatively small
angular radius of $\sim$ 0.35\arcdeg, corresponding to 28 pc at the near kinematic
distance of 4.7 kpc or a more appropriate 67 pc at the far distance.  At the
far distance, the CO luminosity of MC--71 implies a total molecular mass of 4
$\times$ 10$^6$ \sm, about the same as that of the GMC associated with \sgra\ 
(\cite{cor97}) and comparable to the masses of the half dozen or so
largest GMCs in the Galaxy (Dame et al. 1986; Solomon et al. 1987).  

It is worth noting that MC--71 lies within $\sim$ 0.5\arcdeg\ and $\sim$ 15 \kms\ of
another GMC with very similar observational properties (size, mass).  As Figure 2 shows,
the GMC labeled MC--56 has about the same angular size as MC--71, and its
composite linewidth is even larger than that of MC--71; therefore
the far kinematic distance is again strongly favored by the radius-linewidth
relation.  Both these clouds coincide with a remarkable cluster of \hii\ 
regions discussed by Mezger et al. (1970 -- see especially their Fig. 2); they
note that the cluster is very restricted in longitude but has a wide
velocity spread (from $\sim$ --90 to --30 \kms).  It is possible that the abundance
of star formation associated with these two GMCs, including \sgrd, may
have resulted from their collision or tidal interaction.   

The detection of the MC--71 molecular cloud gives a natural explanation for the hydroxyl radical (OH) maser detection
from SNR \gg\ by Frail et al. (1996). As these authors pointed out, this maser line is collisionaly excited by the 
supernova remnant shock going through a molecular cloud.
We note that our distance measurement of 11.0 $\pm$ 0.3 kpc for SNR \gg\ is not consistent
with the value of 5.8 kpc derived by Case \& Bhattacharya (1998) using a new statistical $\Sigma$--D relation.

\section{Distance and nature of \sgrd}

The total hydrogen column density toward the persistent X-ray counterpart of \sgrd\ is
estimated to be (7.7 $\pm$ 0.8) $\times$ 10$^{22}$ cm$^{-2}$ (\cite{woo99b}). Using the empirical relation of 
Predehl \& Schmitt (1995),
this corresponds to an optical extinction of 43.0 $\pm$ 3.9 magnitudes. As the SNR \gg\ is probably at a distance of 
11.0 kpc and associated with MC--71, we will estimate the extinction up to the MC--71 cloud. 
The contribution of each cloud to the extinction is presented in Table 2. We note that the intrinsic extinction 
of MC--71 is very high ($\sim$ 40 magnitudes).
The five molecular clouds with velocity lower than that of MC--71 (MC--122, MC--117, MC--110, MC--97 and MC--83)
must be located in front of it, since both the near and far kinematic distances of these clouds are smaller than
the distance of MC--71.
The two clouds at higher velocity, MC--41 and MC--32, are apparently also in front of MC--71, at their near 
kinematic distances, since hydrogen absorption lines at the velocities of these clouds are seen against the 
radio continuum of the SNR \gg\ (\cite{sar97}).
 
Summing the contributions from all the molecular clouds in front of MC--71 yields a total optical extinction 
of 30.0 $\pm$ 1.9 magnitudes. Contribution to the extinction from atomic hydrogen can be estimated as follows.
Integrating all \hi\ emission from --150 \kms\ to 0 \kms\ gives an upper limit of 11.4 magnitudes.
It is possible that part of the \hi\ emission from --71 \kms\ to 0 \kms\ arises from gas beyond MC--71; by
assuming that half of this \hi\ emission is at the near distance and the other half at the far distance, 
we can deduce a lower limit of 8.0 magnitudes. Therefore, atomic hydrogen adds another 9.7 $\pm$ 1.7 magnitudes,
giving a total extinction to MC--71 of 39.7 $\pm$ 2.5 magnitudes, in good agreement with the value 43.0 $\pm$ 3.9 mag. 
determined for the X-ray counterpart.
Since MC--71 itself would increase the total extinction to 81.1 $\pm$ 4.5 magnitudes, a factor of two above 
the value derived for SAX~J1635.8--4736, \sgrd\ must be located on the near side of MC--71.
The distance to this cloud is obviously the same as that of the CTB~33 complex, i.e. \sgrd\ is at 
distance from the sun of 11.0 $\pm$ 0.3 kpc. This constitutes the second accurate distance estimate of a SGR. 
We should note that the other SGR for which a reasonable distance estimate exists -- \sgra\ at 14.5 $\pm$  1.4 kpc 
(\cite{cor97}) -- is also on the edge of a very massive GMC. This might indicate that these objects
are still lying close to their birth sites. 

The first detailed observations of \gg\ by MOST (\cite{whi96}) showed a complex morphology which, as 
mentioned above, has been resolved into various components by Sarma et al. (1997). Their observations clearly 
revealed the intrinsic morphology of this supernova remnant (Fig. 3 in Sarma et al. (1997)), and therefore provides
the best available radio map of \gg. Therefore, the SNR map of Sarma et al. (1997) is more appropriate for the 
study of \gg\ than the much more widely used MOST map.

The angular size (radius of $\sim$ 45\arcsec) of SNR \gg\ implies a radius of 2.4 pc at a distance of 11.0 kpc.
Following Kafatos et al. (1980), who have studied the expansion of SNRs in various environments,
the size of \gg\ is not in agreement with an expansion into a molecular cloud but is rather typical of a
very young SNR ($<$ 1,000 years) expanding into the interstellar medium. 
This is consistent with our conclusion that \gg\ is on the outer edge of MC--71 and with the fact that SGRs 
are believed to be associated with young SNR.

The persistent X-ray counterpart (the center of its associated error box (\cite{hur99b})) is located 105 $\pm$ 26 \arcsec\ 
away from the 
projected center of the SNR \gg, which corresponds to a displacement of 5.4 pc with our distance estimate of \sgrd.
A very young SNR with an age of 1,000 years would require an unrealistic transverse velocity of 5,400 $\pm$ 1,300 \kms\ 
to reach
this displacement, while a $\sim$ 5,000 years old SNR would imply a more reasonable velocity of 1,080 $\pm$ 270 \kms\ 
(requiring the expansion of the SNR into a {\it rarefied} interstellar medium). 
The latter number seems more likely as high velocity neutron stars have been found with transverse velocities 
up to $\sim$ 1,000 \kms\ (\cite{har93}).  
Therefore it is possible that \sgrd\ is a high velocity neutron star escaping the young SNR \gg.
These numbers could be refined with an improved position of the X-ray counterpart.

\section{Conclusions}

\sgrd\ is found to be located on the edge of a very massive GMC at a distance of 11.0 $\pm$ 
0.3 kpc, with an optical extinction of $\sim$ 43 magnitudes. 
This is the second SGR located on the outer edge of a very large and massive GMC. 
We present evidence for an association with the supernova remnant \gg, which is interacting with the GMC. 
The position of \sgrd\ relative to the SNR indicates that \sgrd\ is probably 
escaping the young ($\sim$ 5,000 years) supernova remnant \gg\ with a high transverse velocity ($\sim$ 1,000 \kms).

\acknowledgements
S.C. would like to thank I. Grenier for useful discussions on the X$_{CO}$ factor.
This research has made use of the SIMBAD database, operated at CDS, Strasbourg, France, and of the Astronomical 
Data Center (ADC) at NASA Goddard Space Flight Center.

\newpage 

\figcaption[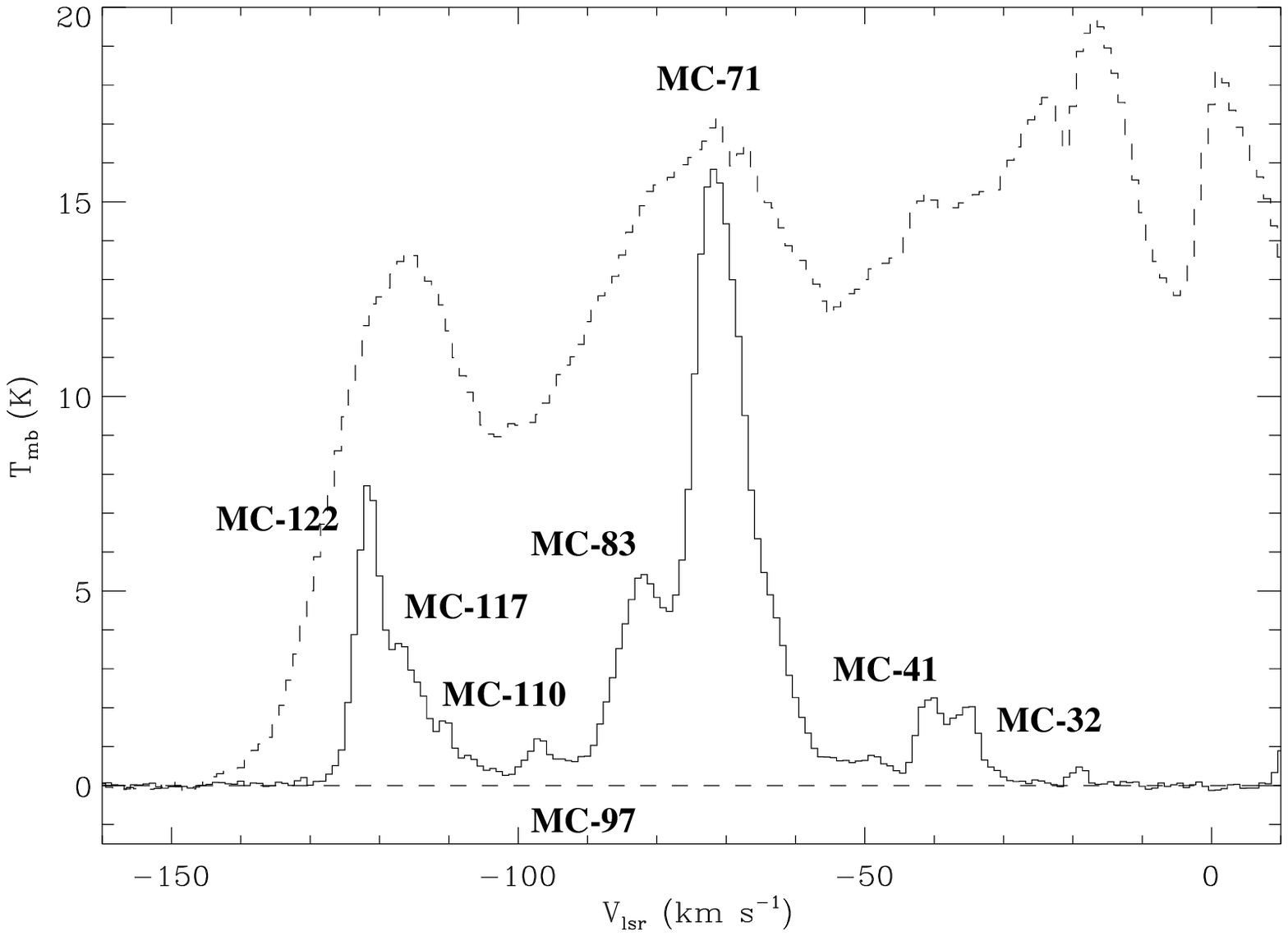]{$^{12}$CO(J=2--1) spectrum toward \sgrd. The different molecular clouds are labeled.
The dotted line represents the \hi\ spectrum at l=337.\arcdeg 0 and b=-0.\arcdeg 125 (intensity divided by 4.5).
Positive velocities \hi\ emission arises from gas beyond the solar circle.
Antenna temperature have been converted  into main beam brightness temperature. }

\figcaption[CORBEL_SGR1627_Fig3.ps]{Map of the CO emission, integrated between the velocities --80 and --40 \kms, 
from the survey of Bronfman et al. (1989). The CO gray levels are spaced at 12 K \kms, starting at 30 K \kms. 
Two GMCs are detected with central velocities of --71 (MC--71) and --56 \kms\ (MC--56).
Contours indicate radio continuum emission at 4850 MHz (Condon et al. 1994), starting at 2.4 Jy beam$^{-1}$ and 
spaced by 2.4 Jy beam$^{-1}$. 
The positions of the persistent X-ray counterpart to \sgrd, the SNR \gg\ and the \hii\ regions G~337.1--0.2 and 
G~336.8+0.0 are indicated.}

\newpage

\begin{deluxetable}{cccccc}
\scriptsize
\tablecaption{Derived parameters from the $^{12}$CO(J=1--0) and $^{12}$CO(J=2--1) spectra for each of the molecular clouds 
along the line of sight to \sgrd.}
\label{table1}
\tablecomments{* We did not attempt to resolve the distance ambiguity for MC-122, MC--117, MC--110, MC--97 and
MC--83, since they are in front of MC--71 in either case.}
\tablewidth{0pc}
\tablehead{
\colhead{Name} & \colhead{V$_{lsr}$} & \colhead{$\Delta$V(FWHM)} & \colhead{Near Distance} &
\colhead{Far Distance} & \colhead{Estimated Distance} \\
\colhead{}              & \colhead{(km s$^{-1}$)} &  \colhead{(km s$^{-1}$)} & \colhead{(kpc)} &
\colhead{(kpc)} & \colhead{(kpc)} \\
}
\startdata
MC--122 & -121.5  & 4.6  & 6.7 & 9.0  &   *  \nl
MC--117 & -116.6  & 5.5  & 6.4 & 9.2  &   *  \nl
MC--110 & -110.3  & 4.0  & 6.1 & 9.5  &   *  \nl
MC--97  & -96.5   & 4.0  & 5.6 & 10.1 &   *  \nl
MC--83  & -83.3   & 5.0  & 5.1 & 10.6 &   *  \nl
MC--71  & -70.9   & 11.0 & 4.6 & 11.0 & 11.0 \nl
MC--41  & -40.6   & 7.0  & 3.1 & 12.5 & 3.1  \nl
MC--32  & -32.2   & 5.7  & 2.6 & 13.0 & 2.6  \nl
\enddata
\end{deluxetable}

\begin{deluxetable}{cccc}
\scriptsize
\tablecaption{Contribution to the molecular hydrogen column density and to the optical extinction of each 
of the molecular clouds along the line of sight to \sgrd.}
\label{table2}
\tablewidth{0pc}
\tablehead{
\colhead{Name} & \colhead{W(CO)}  & \colhead{N(H$_{2}$)}            & \colhead{\av} \\
\colhead{}     & \colhead{(K km s$^{-1}$)} & \colhead{(10$^{21}$ cm$^{-2}$)} &  \colhead{(mag.)}
}
\startdata
MC--122 &  32.9 $\pm$ 3.3 & 6.3 $\pm$ 0.9 & 7.0 $\pm$ 1.1 \nl
MC--117 &  29.0 $\pm$ 2.9 & 5.5 $\pm$ 0.8 & 6.2 $\pm$ 0.9 \nl
MC--110 &  11.0 $\pm$ 1.1 & 2.1 $\pm$ 0.3 & 2.3 $\pm$ 0.4 \nl
MC--97  &  6.1  $\pm$ 1.2 & 1.2 $\pm$ 0.2 & 1.3 $\pm$ 0.3 \nl
MC--83  &  28.3 $\pm$ 2.9 & 5.4 $\pm$ 0.8 & 6.0 $\pm$ 0.9 \nl
MC--71  &  187.1$\pm$ 14.0& 35.5$\pm$ 3.6 & 39.7$\pm$ 4.0 \nl
MC--41  &  20.3 $\pm$ 2.0 & 3.9 $\pm$ 0.6 & 4.3 $\pm$ 0.6 \nl
MC--32  &  13.6 $\pm$ 1.4 & 2.6 $\pm$ 0.4 & 2.9 $\pm$ 0.4 \nl
\enddata
\end{deluxetable}

\newpage

\begin{figure}
\centerline{ \psfig{file=CORBEL_SGR1627_Fig1.ps,angle=0,width=16cm,clip} }
\label{fig1}
\end{figure}

\begin{figure}
\centerline{ \epsfig{file=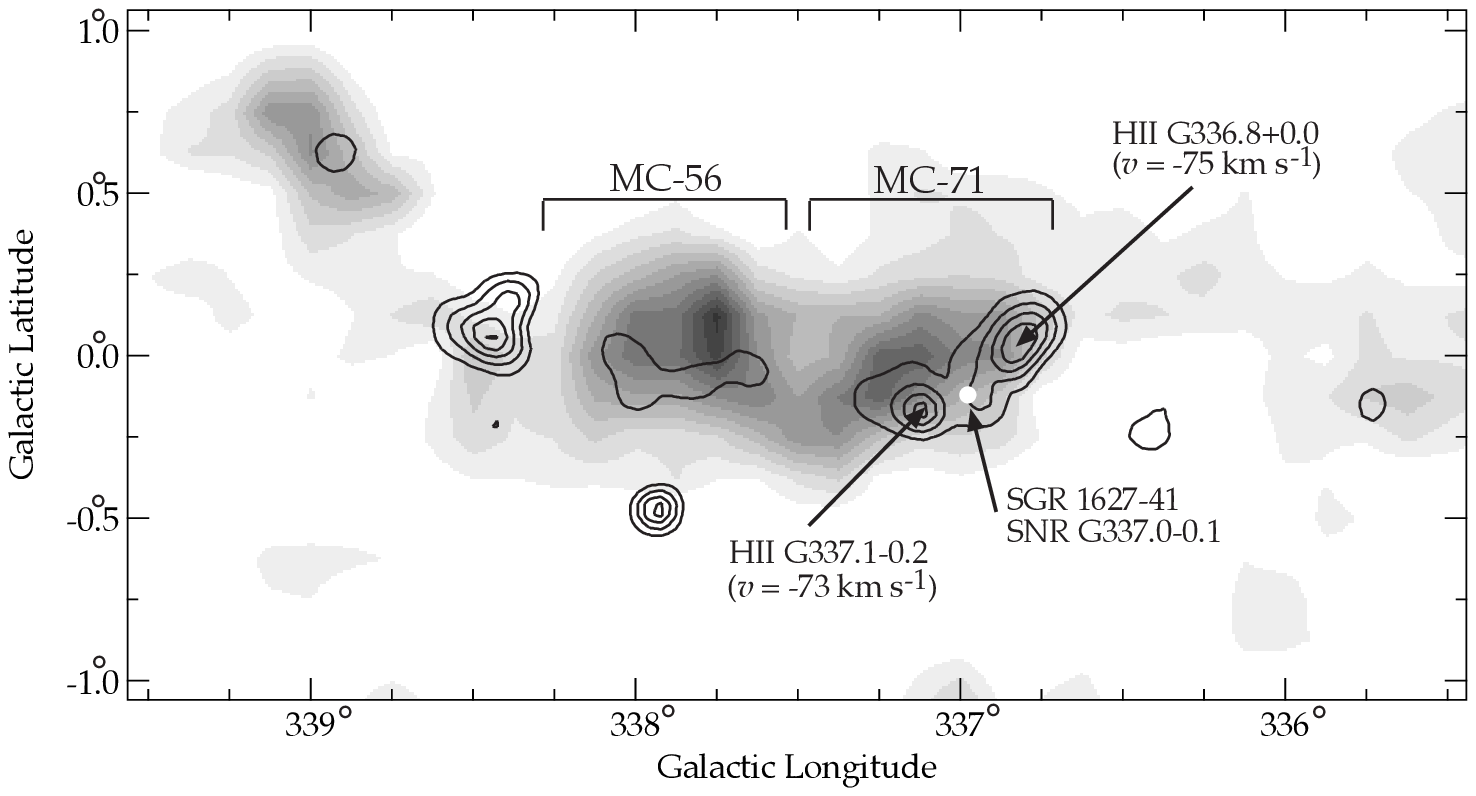,angle=0,width=19cm,clip} }
\label{fig2}
\end{figure}

\end{document}